# Wavelet Analysis for Time Series Financial Signals via Element Analysis

Nathan Zavanelli
*George W. Woodruff School of Mechanical Engineering,*
*College of Engineering, Georgia Institute of Technology*
*Atlanta, GA 30332, USA*
nzavanelli@gatech.edu

*Abstract*— The method of element analysis is proposed here as an alternative to traditional wavelet-based approaches to analyzing perturbations in financial signals by scale. In this method, the processes that generate oscillations in financial signals are modelled as scaled, shifted, and isolated events that produce ripples of various frequencies across a sea of noise as opposed to a simple sinusoidal or mixed frequency oscillation or an impulse. This allows one to directly estimate the wavelet parameters derived only from the generating functions, rejecting spurious perturbations driven by noise or extraneous factors. Financial signals may then be reconstructed based on a finite set of generators localized in time and frequency. This method offers a marked advantage compared to traditional econometric tools because it directly targets the generators of oscillations. Furthermore, the choice of the Morse wavelet allows for wide latitude in capturing a broad set of diverse generators. In this work, the basic mathematical principles underlying element analysis are presented, and the method is applied to the study of variance in financial data, where the advantages of element analysis over traditional wavelet techniques is demonstrated. Specifically, in the example analysis of inflation expectations, element analysis shows a clear ability to distinguish between oscillations formed by noise and those formed by generators logically matched to historical events.

## I. Introduction

Wavelet transforms are a powerful tool for analyzing financial data because they decompose the fluctuations in a signal (like a graph of stock price vs time) into different frequency scales. This multi-resolution analysis is increasingly used to isolate trends by time scale, derive scale-based assessments of data variance, and assess correlation between signals by scale[1-6]. For instance, Crowley et al used a continuous wavelet transform (CWT) to analyze growth cycles of productivity in the European Union (EU), United States (US) and United Kingdom (UK), and they discovered that cycles occurred at various frequencies beyond those classically studied[7,8]. Furthermore, they characterized the correlation between each region's productivity cycles by frequency scale, enabling them to hypothesize how international and national factors drive production volatility in short- and long-term scales. Similar analysis has been conducted for high frequency stock trading, analyzing market trends, assessing relations between variables and the yield curve, and quantifying risk[4,6,9].

However, frequency decomposition techniques, like the wavelet transform, have not achieved their full potential in finance because the mathematical tools have not been sufficiently updated in conjunction with recent discoveries in adjacent fields[10,11]. In order to better understand the problem of frequency decomposition, let us consider the development of suitable approaches from simplest to most complex. The Fourier transform is the simplest frequency decomposition technique, representing a signal as a sum of sinusoidal variations at different frequencies. However, this method is ill-suited for handling non-sinusoidal signals[12]. On the other extreme, the modulus maxima method can be used to analyze signals that are nearly impulses[13,14]. However, almost all financial signals fall at neither extreme, instead exhibiting complex morphologies positioned over a background of noise[15]. These morphologies are well represented by a series of events localized in time with varying spatial distributions and oscillatory and non-oscillatory components[15]. Thus, an effective means for studying these signals is to model them as a sum of various scaled orthogonal wavelets, or the wavelet transform[2,4,16]. However, this transform does not sufficiently separate signal from noise for two reasons. First, any waveform component, be it noise or signal, is mapped to a wavelet scale without any means of distinguishing the two. Second, the signal almost always does not exactly match the chosen wavelet, so it is itself dispersed across several scales. The result is a blurred transform, where significant information may be lost due to the presence of noise[4]. Several traditional methods are commonly used to address this issue, like wavelet thresholding and complex statistical tests[6,10,11]. These approaches, however, are also limited. In the first case, statistically significant wavelet coefficients are identified and maximized, but the underlying limitations of the wavelet transform are never addressed[8,12]. In the second, one typically must make strong assumptions about either the duration or form of a signal, which can lead to significant biases in analysis and great difficulty in application[10].

Instead, a new method termed element analysis, developed by Lilly, can produce a much clearer distinction between signal and noise[16]. The key intuition is to model the processes that generate perturbations in financial signals as scaled, shifted, and isolated events that produce ripples of various frequencies across a sea of noise as opposed to a simple sinusoidal or mixed frequency oscillation or an impulse. Here, a time series signal $x(t)$ is modelled not as a sum of sine waves, impulses, or wavelets, but instead as a baseline of stationary and Gaussian noise upon which are added many individual copies of a complex valued function $\Psi(t)$ with a morphology and time localization that is simply controlled by a time-offset, phase shift, and scaling.

$$x(t) = \sum_{n=1}^{n} \Re\left\{c_n \Psi_{\mu,\gamma}\left(\frac{t-t_n}{\rho_n}\right)\right\} + x_e(t) \quad 1.1$$

where the complex parameter $c_n = |c_n|e^{i\phi_n}$ sets the amplitude $|c_n|$ and phase $\phi_n$ of the event $t_n$ and $\rho_n$ sets the event scale. $x_e(t)$ represents the aforementioned noise. This representation (1.1) is referred to as the element model. Element analysis based on this model is similar to the CWT, but it limits the signal reconstruction only to isolated points in both time and frequency that correspond to specific events, rejecting spurious noise. In general, this method has three steps. First, the wavelet transform maxima corresponding only to events are identified. Second, the significant of these maxima is examined in relation to the noise threshold. Third, the reconstruction is performed based on the coefficients resulting from these maxima. Element analysis is a distinct improvement over wavelet analysis because its goal is not to faithfully capture all signal content, like the CWT, but instead to infer properties of key signal events over a noise threshold. In essence, element analysis seeks to assess the significance of signal events over the null hypothesis of white noise. This method allows for a clear distinction of financial signals separate from the noise, marking a strong improvement over traditional wavelet approaches. Although element analysis has been successfully employed for a variety of signal processing disciplines, it has not been employed for econometrics to the author's knowledge, marking a large missed opportunity in financial data analysis[17,18].

The remainder of the paper will consist of the following sections: a brief discussion of essential wavelet principles, a general summary of the element method, an example relating to financial volatility analysis, and a discussion. In conjunction with his seminal paper, Lilly created a freely available toolbox of Matlab functions, called jLab, available at http://www.jmlilly.net[16]. Furthermore, all software and data relating to the econometrics techniques discussed here is made available by the author at https://github.com/nzavanelli/Element_Analysis_Financial_Data

## II. WAVELET ESSENTIALS

This section seeks to briefly cover several of the key wavelet properties needs to understand element analysis. For further details, please see the following references. These next two sections will also represent a simplification of the material presented in Lilly's work, which the reader may also reference[16]. This section is divided into 2 subsections: (a) continuous wavelet transforms based on the Morse wavelet and (b) additional Morse wavelet properties.

### A. CWT approaches with the Morse Wavelet

The Morse wavelet $\Psi_{\beta,\gamma}$ is a complex function represented for $\beta \geq 0 \ and \ \gamma > 0$ as follows:

$$\Psi_{\beta,\gamma} = \alpha_{\beta,\gamma}\omega^\beta e^{-\omega^\gamma} \times \begin{cases} 1 & \omega > 0 \\ \frac{1}{2} & \omega = 0 \\ 0 & \omega < 0 \end{cases} \quad 2.1$$

where $\beta$ is the order, which controls the low frequency behavior, $\gamma$ the family, controlling the high frequency decay, $\omega$ the frequency, and $\alpha_{\beta,\gamma}$ the normalizing constant of

$$\alpha_{\beta,\gamma} = 2\left(\frac{e\gamma}{\beta}\right)^{\frac{\beta}{\gamma}} \quad 2.2$$

With this definition, the Morse wavelet is strictly analytic, meaning that it must contain both complex and real components. Therefore, the wavelets may be naturally grouped into odd and even pairs, allowing them to capture phase information similar to sine and cosine representations. The wavelet transform of a signal $x(t)$ is represented in the time domain and frequency domain, respectively, as follows:

$$W_{\beta,\gamma}(\tau,s) = \int_{-\infty}^{\infty} \frac{1}{s} \Psi^*_{\beta,\gamma}\left(\frac{t-\tau}{s}\right) x(t)dt$$
$$= \frac{1}{2\pi}\int_{-\infty}^{\infty} e^{i\pi\tau}\Psi^*_{\beta,\gamma}(s,\omega)X(\omega)d\omega \quad 2.3$$

where $X(\omega)$ denotes the Fourier transform of x(t) defined as

$$x(t) = \frac{1}{2\pi}\int_{-\infty}^{\infty} e^{i\pi\tau}X(\omega)d\omega \quad 2.4$$

This transform in the time domain is simply the inner product of the signal and shifted, time scaled versions of the Morse wavelet. In the frequency domain, the scale variable s represents the stretching or compression of the signal, and the rescaled frequency domain wavelet will always be maximized at $\omega_s = \frac{\omega_{\beta,\gamma}}{s}$, which is referred to as the scale frequency. Note that normalization by $\frac{1}{\sqrt{s}}$ is typically performed to ensure the wavelet maintains constant energy. However, $\frac{1}{s}$ normalization is employed here because it allows for the transform values to be controlled by only $c_n$ and not $\rho_n$, greatly simplifying the analytic calculations employed in element analysis.

## B. Morse wavelet properties

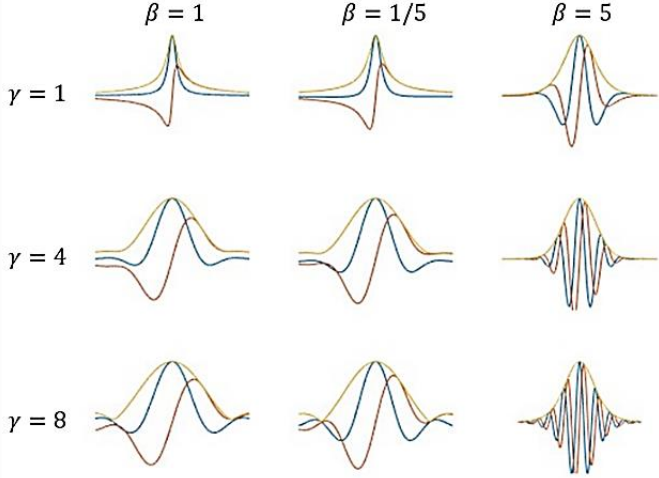

**Figure 1. Morse wavelet representations with various $\beta$ and $\gamma$ values.** Here, the real, imaginary, and envelope components are illustrated as blue, red, and yellow, respectively.

One highly attractive feature of Morse wavelets is that they can assume a wide range of morphologies, which is easily controlled by the choice of $\beta$ and $\gamma$. This is illustrated in Fig 1. Increasing $\beta$ tends to make the signal more oscillatory, and increasing $\beta$ with a fixed $\gamma$ causes more oscillations to fit in the same envelope. On the other hand, modifying $\gamma$ tends to modulate the overall function and envelope shape.

## III. ELEMENT ANALYSIS

This section pertains to the method of element analysis developed by Lilly, representing a summary treatment[16]. Here, the Morse wavelet is introduced as a signal element in 3(a). Next, it is shown in 3(b) that a wavelet transform of a Morse function is in fact another Morse wavelet. This allows for the derivation in 3(c) of the element analysis method to produce transform maxima. Finally, the algorithm is completed in 3(d) by reproducing the element properties based on these maxima.

### A. Morse wavelet representations of signal elements

Consider the wavelet function in (1.1), where $\mu$ and $\gamma$ determine the element function properties (as described in Fig 1) and $\rho$ serves as the scale $s$. Taking the wavelet transform of (1.1) with a Morse wavelet $\Psi^*_{\beta,\gamma}\left(\frac{t-\tau}{s}\right)$ leads to

$$W_{\beta,\gamma}(\tau,s) = \frac{1}{2}\sum_{n=1}^{n} c_n \int_{-\infty}^{\infty} \frac{1}{s} \Psi^*_{\beta,\gamma}\left(\frac{t-\tau}{s}\right) \Psi_{\mu,\gamma}\left(\frac{t-\tau}{\rho_n}\right) dt + \varepsilon_{\beta,\gamma}(\tau,s) \quad 3.1$$

where $\varepsilon_{\beta,\gamma}(\tau,s)$ represents the wavelet transform of the noise process in (1.1). Now, let us define the wavelet maxima points as the time and scale coordinates where the wavelet transform modulus is maximized. This will occur when the following four conditions are met.

$$\frac{\partial}{\partial \tau}\left|w_{\beta,\gamma}(\tau,s)\right| = 0 \quad 3.2$$

$$\frac{\partial}{\partial s}\left|w_{\beta,\gamma}(\tau,s)\right| = 0 \quad 3.3$$

$$\frac{\partial^2}{\partial t^2}\left|w_{\beta,\gamma}(\tau,s)\right| < 0 \quad 3.4$$

$$\frac{\partial^2}{\partial s^2}\left|w_{\beta,\gamma}(\tau,s)\right| < 0 \quad 3.5$$

The goal of element analysis is to simply to use the values of the wavelet transform at these maxima points to estimate the coefficients $c_n$, the scales $\rho_n$, and the times $t_n$ of the N signal events that constitute the signal. From there, a highly denoised scalogram containing only the event content may be produced.

### B. Wavelet transform of a Morse function

When one performs a wavelet transform of a $\mu$ order Morse wavelet $\Psi_{\mu,\gamma}\left(\frac{t}{\rho}\right)$ with a $\beta$ order wavelet of the same family $\gamma$, the result is a modified wavelet $\zeta_{(\beta,\mu,\gamma)}\left(\frac{\tau}{\rho},\frac{s}{\rho}\right)$, as shown in 3.1. This transform is defined as

$$\zeta_{(\beta,\mu,\gamma)}(\tau,s) = \frac{\alpha_{\beta,\gamma}\alpha_{\mu,\gamma}}{\alpha_{\beta+\mu,\gamma}} \frac{s^\beta}{\left(\sqrt[\gamma]{s^\gamma+1}\right)^{\beta+\mu+1}} \Psi_{\beta+\mu,\gamma}\left(\frac{\tau}{\sqrt[\gamma]{s^\gamma+1}}\right) \quad 3.6$$

For a rigorous derivation, please refer to Lilly's work. Briefly, this result may be obtained by substituting the wavelet definition, evaluating the triple integral, rescaling the wavelet, and performing a simple change of variables. Notably, 3.6 shows that performing a wavelet transform of a Morse wavelet modifies the time and scale of the original wavelet, but does not affect the transform amplitude. The result is a wavelet of order $\beta + \mu$, which follows because both $\beta$ and $\mu$ are powers of $\omega$ in the frequency domain, where the wavelet transform corresponds to multiplication.

This modified wavelet also has two intriguing properties: First, the amplitude of the wavelet transform is highly dependent on the scales $s$ and $\rho$. Second, the wavelet's time argument can be effectively rescaled by the transform scale $s$. To examine the scaling effect on 3.6 in more detail, consider the following two cases:

$$\zeta_{(\beta,\mu,\gamma)}(\tau,s) = \frac{\alpha_{\beta,\gamma}\alpha_{\mu,\gamma}}{\alpha_{\beta+\mu,\gamma}} \times \begin{cases} \frac{\rho^{\mu+1}}{s} \Psi_{\beta+\mu,\gamma}\left(\frac{t}{s}\right) & s \gg \rho \\ \frac{s^\beta}{\rho} \Psi_{\beta+\mu,\gamma}\left(\frac{t}{\rho}\right) & s \ll \rho \end{cases} \quad 3.7$$

The result is that when $s \gg \rho$, the resultant wavelet $\zeta_{(\beta,\mu,\gamma)}(\tau,s)$ is smoothed, with the transform spread out over the scale $s$. This is because the transform wavelet $\Psi_{\beta,\gamma}\left(\frac{t}{s}\right)$ is much broader than the

Morse wavelet being transformed $\Psi_{\mu,\gamma}\left(\frac{t}{\rho}\right)$. In the opposite case, the wavelet scale becomes fixed at $\rho$, decreasing in magnitude with further decreases in s.

## C. Transform maxima

The modified wavelet function $\zeta_{(\beta,\mu,\gamma)}(\tau,s)$ can be used to identify the wavelet transform values at the maxima. First, consider the wavelet transform definition from 3.1 with the modified wavelet:

$$W_{\beta,\gamma}(\tau,s) = \frac{1}{2}\sum_{n=1}^{n} c_n \zeta_{(\beta,\mu,\gamma)}\left(\frac{\tau-t_n}{\rho_n},\frac{s}{\rho_n}\right) + \varepsilon_{\beta,\gamma}(\tau,s) \qquad 3.8$$

The expected value of the squared modulus of this wavelet transform thus may be approximated as:

$$E\left\{|W_{\beta,\gamma}(\tau,s)|^2\right\} \approx \frac{1}{4}\sum_{n=1}^{n}|c_n|^2 \left|\zeta_{(\beta,\mu,\gamma)}\left(\frac{\tau-t_n}{\rho_n},\frac{s}{\rho_n}\right)\right|^2 + E\left\{|\varepsilon_{\beta,\gamma}(\tau,s)|^2\right\} \qquad 3.9$$

This approximation requires the assumption that cross-terms within the summation may be neglected on the basis of the zero mean and that events are well separated. The second assumption is generally not fully valid for financial signals, and the result is the potential for low level maxima amplitudes to arise. Fortunately, these amplitudes are generally higher than the noise floor, but still lesser than a pure signal with only one generating function. However, care must be taken when selecting $\beta$ and $\gamma$ parameters to ensure strong maxima in the case of most signals. In general, financial signals with many complicated interactions should avoid large $\gamma$ and small $\beta$ values to ensure strong monotonic decay and avoid sidelobe maxima effects.

Now let us consider the scale locations and wavelet transform values corresponding to the wavelet maxima. Note that the maxima of $\zeta_{(\beta,\mu,\gamma)}\left(\frac{\tau}{\rho},\frac{s}{\rho}\right)$ with respect to time occurs at $\tau = 0$, at which point $\zeta_{(\beta,\mu,\gamma)}\left(0,\frac{s}{\rho}\right)$ assumes the real and positive value:

$$\zeta_{(\beta,\mu,\gamma)}\left(0,\frac{s}{\rho}\right) = \frac{\alpha_{\beta,\gamma}\alpha_{\mu,\gamma}}{2\pi\gamma}\Gamma\left(\frac{\beta+\mu+1}{\gamma}\right)\frac{\left(\frac{s}{\rho}\right)^\gamma}{\sqrt[\gamma]{\left(\left(\frac{s}{\rho}\right)^\gamma+1\right)^{\beta+\mu+1}}} \qquad 3.10$$

This value may be derived by combining 3.6 with the definition of the wavelet function at $\tau = 0$. Defining $\tilde{s} \equiv \frac{s}{\rho}$ and differentiating $\zeta_{(\beta,\mu,\gamma)}(0,\tilde{s})$ with this new variable allows for one to determine that the maximal value occurs at:

$$\zeta_{(\beta,\mu,\gamma)max} = \zeta_{(\beta,\mu,\gamma)}(0,\tilde{s}_{max}) \qquad 3.11$$

$$\tilde{s}_{max} = \left(\frac{\beta}{\mu+1}\right)^{\frac{1}{\gamma}} \qquad 3.12$$

Inserting 3.11 into 3.10 allows for the determination of the maximum value of the modified wavelet transform:

$$\zeta_{(\beta,\mu,\gamma)max} = \frac{\alpha_{\beta,\gamma}\alpha_{\mu,\gamma}}{2\pi\gamma}\Gamma\left(\frac{\beta+\mu+1}{\gamma}\right)\eta_{\beta,\mu,\gamma} \qquad 3.13$$

where $\eta_{\beta,\mu,\gamma}$ is the scale weighting function defined as:

$$\eta_{\beta,\mu,\gamma} \equiv \frac{\tilde{s}_{max}^\gamma}{\sqrt[\gamma]{(\tilde{s}_{max}^\gamma+1)^{\beta+\mu+1}}} = \frac{\left(\frac{\beta}{\mu+1}\right)^{\frac{\beta}{\gamma}}}{\left(\frac{\beta}{\mu+1}+1\right)^{\frac{\beta+\mu+1}{\gamma}}} \qquad 3.14$$

Thus, is can readily seen that the maximum value $\zeta_{(\beta,\mu,\gamma)max}$ is indeed independent of the scale $\rho$.

## D. Estimating element properties from the transform maxima

In the case of well-behaved signals with a proper choice of wavelet parameters, we will have one maximum point for each of the N generating events, and the $n^{th}$ maxima will be located at time $t_n$ and scale $s_n = \rho_n \tilde{s}_{max}$. It is clear from 3.8 that the wavelet transform here is thus:

$$W_{\beta,\gamma}(t_n,s_n) = \frac{1}{2}c_n\zeta_{(\beta,\mu,\gamma)max} \qquad 3.15$$

Now, one may use the equations in 3.2-3.5 to define the set of observed time/scale maxima points, which will be denoted as $(\hat{t}_n,\hat{s}_n)$. From these points, the element properties $(t_n,c_n,\rho_n)$ may be simply estimated. If one defines $W_n \equiv W_{\beta,\gamma}(\hat{t}_n,\hat{s}_n)$ as the wavelet transform at each observed maximum, then these element properties become:

$$\hat{t}_n = \hat{t}_n \qquad \hat{c}_n = 2\frac{W_n}{\zeta_{(\beta,\mu,\gamma)max}} \qquad \hat{\rho}_n = \frac{\hat{s}_n}{\tilde{s}_{max}} \qquad 3.16$$

where the quantities are hatted to show that these values are estimates of the true element properties. Before the method is complete, one final modification is necessary. Here, the frequency of the function is reported instead of the scale. This may be rectified by substituting $s = \frac{\omega_{\beta,\gamma}}{\omega_s}$ and $\rho = \frac{\omega_{\mu,\gamma}}{\omega_s}$ into $\hat{\rho}_n = \frac{\hat{s}_n}{\tilde{s}_{max}}$ from 3.16 to yield:

$$\omega_{\hat{\rho}_n} = \omega_{\hat{s}_n}\frac{\omega_{\mu,\gamma}}{\omega_{\beta,\gamma}}\tilde{s}_{max} = \omega_{\hat{s}_n}\frac{\omega_{\mu,\gamma}}{\omega_{\beta,\gamma}}\left(\frac{\beta}{\mu+1}\right)^{\frac{1}{\gamma}} \qquad 3.17$$

which is the relationship between the frequency band $\omega_{\hat{s}_n}$ of the observed wavelet maximum and that of the corresponding element, $\omega_{\hat{\rho}_n}$. We now have all the parameters necessary to reconstruct the signal transform as a scalogram containing the information of the N elements, without the noise function. An algorithm to do so with examples and all code used in this work is proved by the author at https://github.com/nzavanelli/Element_Analysis_Financial_D

ata. Furthermore, the reader is encouraged to consider the original algorithms derived by Lilly et al, which are available at http://www.jmlilly.net and upon which the author's algorithms are heavily based.

## IV. APPLICATION TO VARIANCE ANALYSIS

As mentioned in the introduction, wavelet analysis is a powerful, yet underutilized, tool in econometrics for analyzing financial data by time scale. Although many complex analyzes are possible, like assessing the correlation of variables to the yield curve by scale, two very simple examples will be shown here to demonstrate that the element method of wavelet analysis offers a notable improvement over traditional wavelet methods.

First, let us consider the expected 10 year inflation rate in the United States (E10YRI) between July 2018 and July 2022. Fig 2(A) shows the E10YRI versus time over the period described. A third order high-pass Butterworth infinite impulse response filter with a cutoff of 1/3 years is then applied to isolate only the higher frequency perturbations in the signal, removing any longer-term trends. The result is the graph in Fig 2(B). Next, a traditional wavelet scalogram is produced from the data in Fig 2(B) using a Morse wavelet with parameters $\beta = 3$ and $\gamma = 1$. The resultant scalogram is shown in Fig 2(C). The scalogram appears to show a clear and persistent long-term volatility on the order of multiple months ($1/12 - 1/20$ years), which generally waxes and wanes with time. Furthermore, several shocks are present in March 2020, 2021, and 2022. Interestingly, the volatility associated with the shock in 2022 appears to be notably higher frequency than that of 2021.

The data can be extracted from each of these frequency bands and analyzed separately for a variety of purposes. Although this scalogram does offer a promising and useful tool for analyzing volatility in the E10YRI, a comparison with an element analysis scalogram will reveal several limitations. The element analysis method described in this paper was used with identical wavelet parameters to produce the scalogram shown in Fig 2(D). From this scalogram, several additional high volatility events that are not evident in Fig 2(C) are clearly visible. For instance, an uptick in volatility associated with the stock market contraction of December 2018 is easily visible, and the volatility in this shock can be clearly delineated by frequency scale. In Fig 2(C), this shock is not readily visible, as it is drowned out by noise and the other generating variabilities. However, a careful inspection of Fig 2(B) shows that this volatility is, in fact, present in December 2018. Whereas the traditional scalogram cannot capture this information, the element analysis method can. Likewise, news regarding currency conflict between the EU and US and anxiety over trade with China fueled volatility in July 2019, which is captured in the element analysis scalogram and not the traditional wavelet scalogram. Finally, the twin events of 2021 and 2022 can be better studied via element analysis. Whereas the traditional wavelet scalogram indicates that the volatility in 2022 is bifurcated into two separate modes, the element analysis produces the more intuitive result that the volatility is in fact continuous. Similarly, the start and end times in Fig 2(C) are generally more spread out than those clearly delineated in Fig 2(D). Finally, it is noteworthy that Fig 2(C) generally lacks much of the changes in high frequency volatility that is present in Fig 2(D). Overall, it should be clear from this example that the element analysis method offers a promising alternative to traditional wavelet scalogram methods for analyzing financial data.

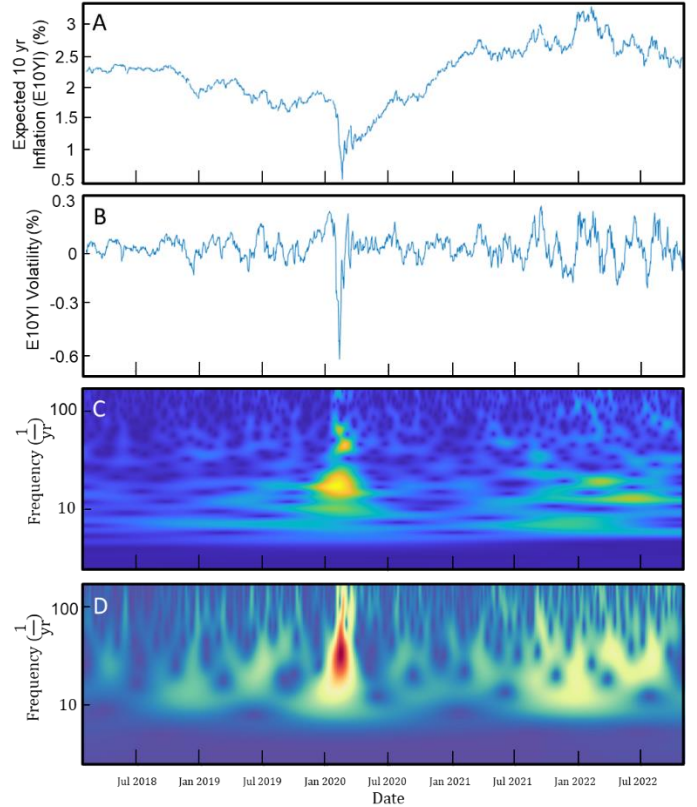

**Figure 2. Wavelet scalogram analysis of expected 10 year inflation rate (E10YRI) in the United States.** (A) Plot of E10YRI vs time. (B) Plot of high frequency filtered perturbations in E10YRI vs time. (C) Wavelet scalogram produced from the data in B. (D) Element analysis scalogram produced from the data in B.

## V. DISCUSSION

Here, the element analysis method of Lilly has been described and, and its application to econometrics has been demonstrated in a simple example[16]. The key intuition is to model the processes that generate perturbations in financial signals as scaled, shifted, and isolated events that produce ripples of various frequencies across a sea of noise as opposed to a simple sinusoidal or mixed frequency oscillation. This method is similar to the continuous wavelet transform and based on the Morse wavelet, but it is unique in that it produces a new transform for each generating event, allowing for a significant improvement in noise reduction and signal clarity. The analysis created by Lilly et al marks a valuable addition to the econometrist's toolbox for analyzing financial signals because it can more precisely capture generators of perturbations in financial signals than traditional wavelet methods. This was demonstrated in an analysis of the expected 10-year inflation rate in the United States (E10YRI) between July 2018 and July 2022, where several clear events were present in the element analysis that could not be studied using traditional wavelet methods. In addition, there are ample opportunities for this method to be improved to better fit financial data. For instance, assumptions regarding the distribution of noise could effect the expected value determined in 3.9, and variance and bias terms could also be introduced at this juncture. There is also a lack of clear criteria to determine the optimal wavelet parameters for the transform wavelet, especially for financial data. Finally, this method could be further generalized by modelling events as superpositions of higher order wavelets as opposed to single functions.